%% file: angdia.tex
\newif\ifREFEREE
 \REFEREEtrue
 \REFEREEfalse

\ifREFEREE
	\documentstyle[epsf,referee]{mn}
\else
	\documentstyle[epsf]{mn}
\fi
	\input mnextra
\newif\ifAMStwofonts
\AMStwofontsfalse

\def\ti{\widetilde}
\def\FL{Friedmann--Lema\^\i tre}
\def\bi#1{{\bmi#1}}
\def\fracj#1#2{{\textstyle{#1\over#2}}}
\newlength{\parwidth}
\newlength{\parindentbak}
\newcounter{example}[section]
\def\beq{\begin{equation}}
\def\eeq{\end{equation}}
\def\beqn{\begin{eqnarray}}
\def\eeqn{\end{eqnarray}}

\renewcommand{\d}{\rmn{d}}

\def\kpc{\mbox{$\,{\rm kpc}$}}

\def\refer#1#2#3{#1, #2, #3}

\def\aj#1#2{\refer{AJ}{#1}{#2}}

\def\apj#1#2{\refer{ApJ}{#1}{#2}}

\def\mn#1#2{\refer{MNRAS}{#1}{#2}}

\def\pasj#1#2{\refer{PASJ}{#1}{#2}}
\def\pasp#1#2{\refer{PASP}{#1}{#2}}

\def\O{\Omega}

\def\a{\alpha}

\def\d{\rmn{d}}
\def\e{\varepsilon}
\def\g{\gamma}
\def\i{\relax\ifmmode{\rm i}\else\char16\fi}

\def\l{\left}
\def\r{\right}
\def\s{\sigma}

\def\dq{\delta q_0}

\def\ba{\bi{\alpha}}

\def\br{\bi{r}}

\def\bxi{\bi{\xi}}

\def\ee{\rmn{e}}

\def\pa{\partial}

\def\mod#1{\l |{#1}\r |}
\def\av#1{\l \langle{#1}\r \rangle}

\def\smallfont{\footnotesize}

\ifoldfss
  \newcommand{\rmn}[1] {{\rm #1}}

  \ifCUPmtlplainloaded \else
    \NewTextAlphabet{textbfit} {cmbxti10} {}
    \NewTextAlphabet{textbfss} {cmssbx10} {}
    \NewMathAlphabet{mathbfit} {cmbxti10} {} 
    \NewMathAlphabet{mathbfss} {cmssbx10} {} 
  \fi
  \ifAMStwofonts
    \ifCUPmtlplainloaded \else
      \NewSymbolFont{upmath} {eurm10}
      \NewSymbolFont{AMSa} {msam10}
      \NewMathSymbol{\upi}     {0}{upmath}{19}
      \NewMathSymbol{\umu}     {0}{upmath}{16}
      \NewMathSymbol{\upartial}{0}{upmath}{40}
      \NewMathSymbol{\leqslant}{3}{AMSa}{36}
      \NewMathSymbol{\geqslant}{3}{AMSa}{3E}

    \fi
  \fi
\fi 

\ifnfssone
  \newmathalphabet{\mathit}
  \addtoversion{normal}{\mathit}{cmr}{m}{it}
  \addtoversion{bold}{\mathit}{cmr}{bx}{it}
  \newcommand{\rmn}[1] {{\rm #1}}

  \newmathalphabet{\mathbfit} 
  \addtoversion{normal}{\mathbfit}{cmr}{bx}{it}
  \addtoversion{bold}{\mathbfit}{cmr}{bx}{it}
  \newmathalphabet{\mathbfss} 
  \addtoversion{normal}{\mathbfss}{cmss}{bx}{n}
  \addtoversion{bold}{\mathbfss}{cmss}{bx}{n}
  \ifAMStwofonts
    \ifCUPmtlplainloaded \else
      \UseAMStwoboldmath
      \makeatletter
      \new@mathgroup\upmath@group
      \define@mathgroup\mv@normal\upmath@group{eur}{m}{n}
      \define@mathgroup\mv@bold\upmath@group{eur}{b}{n}
      \edef\UPM{\hexnumber\upmath@group}
      \new@mathgroup\amsa@group
      \define@mathgroup\mv@normal\amsa@group{msa}{m}{n}
      \define@mathgroup\mv@bold\amsa@group{msa}{m}{n}
      \edef\AMSa{\hexnumber\amsa@group}
      \makeatother
      \mathchardef\upi="0\UPM19
      \mathchardef\umu="0\UPM16
      \mathchardef\upartial="0\UPM40
      \mathchardef\leqslant="3\AMSa36
      \mathchardef\geqslant="3\AMSa3E
    \fi
  \fi
\fi 

\ifnfsstwo
  \newcommand{\rmn}[1] {{\rm #1}}

  \DeclareMathAlphabet{\mathbfit}{OT1}{cmr}{bx}{it}
  \SetMathAlphabet\mathbfit{bold}{OT1}{cmr}{bx}{it}
  \DeclareMathAlphabet{\mathbfss}{OT1}{cmss}{bx}{n}
  \SetMathAlphabet\mathbfss{bold}{OT1}{cmss}{bx}{n}
  \ifAMStwofonts
    \ifCUPmtlplainloaded \else
      \DeclareSymbolFont{UPM}{U}{eur}{m}{n}
      \SetSymbolFont{UPM}{bold}{U}{eur}{b}{n}
      \DeclareSymbolFont{AMSa}{U}{msa}{m}{n}
      \DeclareMathSymbol{\upi}{0}{UPM}{"19}
      \DeclareMathSymbol{\umu}{0}{UPM}{"16}
      \DeclareMathSymbol{\upartial}{0}{UPM}{"40}
      \DeclareMathSymbol{\leqslant}{3}{AMSa}{"36}
      \DeclareMathSymbol{\geqslant}{3}{AMSa}{"3E}
    \fi
  \fi
\fi 

\ifCUPmtlplainloaded \else
  \ifAMStwofonts \else 
    \def\upi{\pi}
    \def\umu{\mu}
    \def\upartial{\partial}
  \fi
\fi

\begin{document}

\title{Gravitational lensing and the angular-diameter distance relation}

\author[Fedja Hadrovi\'c and James Binney]
{	Fedja Hadrovi\'c and James Binney	\\
Theoretical Physics, 1 Keble Road, Oxford OX1 3NP}

\pubyear{1997}

\maketitle

\begin{abstract}
We show that the usual relation between redshift and angular-diameter
distance can be derived by considering light from a source to be
gravitationally lensed by material that lies in the telescope beam as it
passes from source to observer through an otherwise empty universe.  This
derivation yields an equation for the dependence of angular-diameter on
redshift in an inhomogeneous universe. We use this equation to model the
distribution of angular-diameter distance for redshift $z=3$ in a
realistically clustered cosmology. This distribution is such that attempts
to determine $q_0$ from angular-diameter distances will systematically
underestimate $q_0$ by $\sim0.15$, and large samples would be required to
beat down the intrinsic dispersion in measured values of $q_0$.
\end{abstract}


\section{INTRODUCTION}

The large-scale structure of the Universe is believed to be closely
approximated by one of the \FL\ cosmological models. These are characterized
by the values of three parameters: the matter density, the radius of curvature
of spatial sections and the cosmological constant.  Determination of these
values has long been considered one of the fundamental tasks of
observational cosmology. In this connection a potentially key observable is
the relationship between redshift $z$ and the angular-diameter distance
$D(z)$, which is defined to be the ratio of the linear diameter of an object
to the angular diameter that it subtends when observed at redshift $z$.
There have been may attempts to determine $D(z)$
\cite{sandage,kellermann,crawford} which has recently prompted a wider
discussion about the feasibility of the method
\cite{nilsson,dabrowski,kantowski,stephanas}. The form of $D(z)$ depends on
the geometry of the Universe in the sense that the larger the curvature $K$
is, the smaller $D$ is at a given redshift. That is, the more positively
curved the Universe is, the more slowly the angular size of an object
decreases as it is moved away from the observer.

In an inhomogeneous universe deviations of the metric from the \FL\ form
give rise to fluctuations in measured values of $D$ at fixed $z$.  Previous
analyses of this effect \cite{zeldovich,refsdal,dyer-roeder,sasaki,kasai,watanabe}
worked directly from general relativity and produced results of considerable
complexity. By delegating relativistic considerations to the theory of
gravitational lensing \cite{schneider}, we obtain a much simpler analysis
and an equation (\ref{new:distance}) that involves the cosmic density field
rather than the cosmic metric or potential. This simplicity enables us to
evaluate the effects of inhomogeneity for the case of realistic clustering,
rather than the case of either weak perturbations \cite{sasaki} or randomly
distributed point masses \cite{zeldovich,refsdal,dyer-roeder,watanabe,kantowski}.

Over the last decade there has been a growing awareness of the importance of
gravitational lensing for observations of high-redshift objects.
Gravitational lensing and the dependence of $D$ on $z$ are two sides of the
same coin: both phenomena are caused by the tendency of matter that lies
between the observer and a distant object to focus radiation from that
object, thereby increasing its apparent size and brightness. In Section 2 we
demonstrate this connection quantitatively by showing that the standard
formula for $D(z)$ in a \FL\ universe can be obtained by applying
conventional lensing theory to a Universe in which there is matter {\em
only\/} in the telescope beam towards the object under study. In Section 3 we
describe our model of the clustered cosmic density field, and in Section 4
we use this model to calculate probability distributions for $D(z)$ from
objects of various linear sizes. Section 5 sums up.

We throughout use the convention that $\ti D$ and $D$, respectively, denote
angular-diameter distance before and after lensing is taken into account.

Since luminosity distance $D_L$ is rigorously related to angular-diameter
distance by $D_L=D/(1+z)^2$ \cite{ethering}, our distributions of values of
$D$ imply identical distributions of values of $D_L$.

The unit system we use is based on $G=c=H_0=1$, which significantly
simplifies the equations in cosmology.  All lengths quoted are scaled to
$H_0=100 h\, \rmn{km}\, \rmn{s}^{-1}\, \rmn{Mpc}^{-1}$.

\begin{figure}
\epsfxsize=20pc\epsfbox{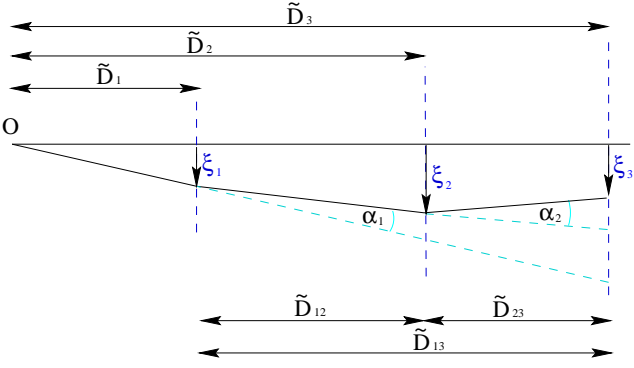}
\caption{A sequence of gravitational lenses}
\end{figure}

\section{The angular-diameter distance relation from lensing}

\subsection{A sequence of gravitational lenses}

Fig.~1 shows a ray that is deflected through angles
$\ba_1$ and $\ba_2$ by two lenses, which it passes at impact parameters
$\bxi_1$ and $\bxi_2$, respectively. From the figure it is immediately
apparent that
\beq
\bxi_3={\ti D_3\over \ti D_1}\bxi_1-\ba(\bxi_1)\ti D_{13}
-\ba(\bxi_2)\ti D_{23}.
\eeq
 The generalization of this equation to an arbitrary number of lenses is
easily seen to be
 \beq\label{lens_seq}
 \bxi_j={{\ti D_j}\over{\ti D_1}}\bxi_1
-\sum_{i=1}^{j-1}{\ti D_{ij}\ba_i(\bxi_i)}.
\eeq

Light that passes at radius vector $\bxi$ through a disc  of matter that has
uniform surface density $\Sigma$ is deflected through an angle
 \beq
\ba=4\pi\Sigma\bxi.
\eeq
 Hence in this case the impact parameters are all parallel and satisfy
\beq\label{all_xis}
 \bxi_j={{\ti D_j}\over{\ti D_1}}\bxi_1
-4\pi\sum_{i=1}^{j-1}\Sigma_i\ti D_{ij}\bxi_i.
\eeq
 Finally, if $\bxi_i$ is a diameter of an object, then the angular-diameter
distance of that object is $D_i\equiv\xi_i/(\xi_1/\ti D_1)$, so on taking the
modulus of (\ref{all_xis}) and dividing through by $(\xi_1/\ti D_1)$ we conclude
that true angular-diameter distances $D_i$ satisfy
\beq\label{disc_ang_diam}
 D_j=\ti D_j-4\pi\sum_{i=1}^{j-1}\Sigma_iD_i\ti D_{ij}
\eeq

\subsection{Application to a homogeneous universe}

We now we show that equation (\ref{disc_ang_diam}) reproduces the familiar
angular-diameter distance equation for a \FL\ universe.  We consider an
empty universe. In such a universe there is nothing to single out a unique
rest frame at any given event, so redshift is not uniquely related to
distance. This permits us simply to adopt the relation $s(z)$ between proper
distance and redshift in a \FL\ universe. We have that $s(z)$
satisfies [Scheneider et al., eq.\ (4.47b)]
 \beq\label{FLsz}
{\d s\over\d z}=(1+z)^{-2}(1+\Omega z)^{-1/2}.
\eeq

From the gravitational-focusing equation [Schneider et al.\ eq.~(3.64)] in
the case of empty space (vanishing Ricci tensor and shear) we have
 \beq\label{focus_e}
{\d^2\ti D\over\d\tau^2}=0,
\eeq
 where $\tau$ is an affine parameter for the light beam.  In terms of the
wavenumber, $k$, we have $\d s/\d\tau\propto k\propto 1+z$, so we may use
equation (\ref{FLsz}) to eliminate $\tau$ from (\ref{focus_e}) in favour of
$s$. We then find that the focusing equation states that in our empty
universe, as a function of $z$, angular-diameter distance $\ti D$ satisfies
 \begin{eqnarray}\label{D_empty1}
(1+z)(1+\O z)&& {\pa^2\ti D(y,z)\over\pa z^2}\nonumber\\
&&+\l(\fracj72\O z+\fracj12\O+3\r){\pa\ti D(y,z)\over\pa z}=0.
 \end{eqnarray}
 We also have the initial condition [Schneider et al.\ eq.~(4.53)]
\beq\label{D_empty2}
{\pa\ti D(y,z)\over\pa z}\bigg|_{y=z}=(1+z)^{-2}(1+\O z)^{-1/2}.
\eeq

Now we fill the telescope beam with the normal matter density of a
\FL\ universe and use equation (\ref{disc_ang_diam}) to
calculate the angular-diameter distance of an object at `redshift' $z$.
We first take the limit of equation (\ref{disc_ang_diam}) in which there are
an infinite number of discs. Since in our units the current critical density
is $3/(8\pi)$, the disc that lies between $z+\d z$ and $z$ has surface
density 
 \beq
\Sigma=(1+z)^3{3\Omega\over8\pi}{\d s\over\d z}\d z,
\eeq
 where $\Omega$ is the usual density parameter. With equation
(\ref{FLsz}) this becomes
 \beq
\Sigma(z)={3\Omega\over8\pi}{1+z\over\sqrt{1+\Omega z}}\d z.
\eeq
 Inserting this expression for $\Sigma$ into equation (\ref{disc_ang_diam})
and proceeding to the limit $\d z\to0$ we find
\beq
D(z)=\ti D(z)
-\fracj32\Omega\int_0^z\d y{1+y\over\sqrt{1+\Omega y}}D(y)\ti D(y,z).
\eeq
 We now convert this integral equation for $D(z)$ into a differential
equation. Differentiating we find
 \begin{eqnarray}
{\d D\over\d z}&=&{\d\ti D\over\d z}-\fracj32\,\O\,
\int_0^z\d y\,{1+y\over\sqrt{1+\O y}}D(y)\l({\pa\over\pa z}\ti D(y,z)\r),
	\nonumber\\
{\d^2D\over\d z^2}&=&{\d^2\ti D\over\d z^2}-\fracj32\,
\O\,{1+z\over\sqrt{1+\O z}}D(z)\l({\pa\over\pa z}\ti D(y,z)\r)_{y=z}\\ 
&&-\fracj32\,\O\,\int_0^z\d y\,{{1+y\over\sqrt{1+\O y}}D(y)
\l({\pa^2\over\pa z^2}\ti D(y,z)\r)}\nonumber.
\end{eqnarray}
 Combining these equations and taking advantage equations (\ref{D_empty1})
and  (\ref{D_empty2}),
we recover the standard equation for the angular-diameter distance in a
conventional \FL\ universe:
 \beq\label{eq:distance}
(1+z)(1+\O z){\pa^2 D\over\pa z^2} + 
\l(\fracj72\O z+ \fracj12\O+3\r){\pa D\over\pa z}+\fracj32\O D=0.
\eeq

\subsection{Application to an inhomogeneous universe}

The most important feature of the above derivation is that it does not
depend on $\O$ being constant. In the first two terms of equation
(\ref{eq:distance}) $\O$ appears as a result of the reparametrisation
($s\mapsto z$).  It is not related to the local matter distribution and can
be thought of as the averaged density parameter $\av{\O}$.
The parameter $\O$ in the last term, ${3 \over 2} \O D$ {\em is\/} related to
the local matter density and comes directly from the gravitational lensing
calculation.  Therefore, in the case of a locally inhomogeneous universe
that approaches a \FL\ model in the large-scale limit, we can
write
 \begin{eqnarray}\label{new:distance}
(1+z)(1+\av{\O} z){\pa^2 D\over\pa z^2} + 
\big(\fracj72\av{\O}z+&\fracj12&\av{\O}+3\big){\pa D\over\pa z}\nonumber\\
&+&\fracj32\O(z) D=0.
\end{eqnarray}
 Note that $\O(z)$ describes the comoving matter density because the
physical density is $\rho(z)=\fracj3{8\pi}\O(1+z)^3$.

Simply replacing $\O$ by $\av{\O}$ in all but the last term of equation
(\ref{eq:distance}) does not allow for a complete discussion of the effects
of inhomogeneity on images: in addition to being magnified by matter within
the beam, images will be distorted and may be even split into multiple
images. We have neglected these potentially important effects by (i)
assuming that the material that lies between redshifts $z+\d z$ and $z$
forms a uniform disc, and (ii) neglecting shear that is induced by clumps of
material that lie outside the beam. Futumase \& Sasaki (1989) and Watanabe
\& Sasaki (1990) show that as long as the scale of inhomogeneities is
greater than, or equal to galactic scale, shear does not contribute
significantly to focusing.  

\begin{figure}
\centerline{\epsfxsize=12pc\epsfbox{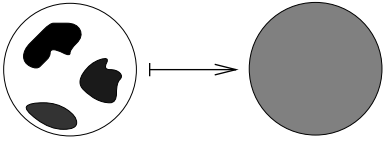}}
\caption{Averaging of the matter distribution over the beam cross section}
\label{fig:average}
\end{figure}

By contrast, the assumption that the beam is filled by a series of
uniform-density discs constitutes a non-trivial approximation about the
matter distribution in the beam, namely that we may average the density
across the beam  as shown in Fig.~\ref{fig:average}.

\section{Statistical model of the  field $\O(\br)$}
\label{sec:matter}

We now investigate the predictions of the generalized diameter-distance
equation (\ref{new:distance}). For this investigation we require a
statistical description of the density field along the telescope beam.
This is a random field, which we think of as a function of comoving distance
$x$. We assume that $\Omega(x)$ follows a log-normal distribution -- see
Coles \& Jones (1991) for a discussion of the characteristics and advantages
of the log-normal distribution in cosmology.  We confine ourselves to the
case of a critical-density universe: $\av{\O}=1$. With these assumptions
$\O(x)$ is given by 
 \beq\label{Ofrome}
\O(x)={\ee^{\e(x)}\over\av{\ee^\e}},
\eeq
 where $\e(x)$ is a Gaussian random field. Without loss of generality we
set $\av{\e}=0$.

We define the  two-point correlation function, $\xi_f$ of a field $f(x)$
by 
 \beq
\xi_f(x)={\av{f(x'+x)f(x')}-\av{f(x')}^2
\over\av{f(x')^2}-\av{f(x')}^2}.
 \eeq
The correlation functions of the fields $\O(x)$ and $\e(x)$ are related by
 \beq\label{xiO:xie}
\xi_\O(x)={\exp(\s_\e^2\xi_\e(x))-1\over\exp(\s_\e^2)-1},
\eeq
 where $\sigma_\e^2=\av{\e^2}$ is the variance of the Gaussian field.

The Gaussian field $\e(x)$ is determined by its power spectrum
$P_\e(k)$, which is essentially the Fourier transform of $\xi_\e$:
 \beq\label{Pe:xie}
P_\e(k)={\sigma_\e^2\over2\pi}\int\d x\,\ee^{\i kx} \xi_\e (x).
\eeq
 Hence, if we know $\xi_\O(x)$, we may construct realizations of $\O$ by
determining $\xi_\e(x)$ from equation (\ref{xiO:xie}) and then using
equation (\ref{Pe:xie}) to determine $P_\e(k)$.

The galaxy correlation function may be
approximated by \cite{padmanabhan}
 \beq
\xi(r)=\l({r \over {r_c}}\r)^{-\g},
\eeq
 where $\g\approx1.8$ and the correlation length is
$r_c\simeq5.5h^{-1}\rmn{Mpc}$.  The correlation function of the density
field is often assumed to have the same form, but a different amplitude.
The bias factor is introduced by setting
 \beq
b={\xi_{\hbox{\scriptsize{\sc galaxies}}}\over\xi_{\hbox{\scriptsize{\sc
matter}}}}. 
\eeq
 Measurements indicate that $1\la b \la 2$.  Hence we require $\xi_\O$ such
that $\xi_\O(0)=1$ and 
 \beq
 \s_\O^2\xi_\O(r)\approx b^{-1}\l({r \over {r_c}}\r)^{-\g}.
\eeq
 We have adopted the form
 \beq\label{xiofO}
\xi_\O(r)=\l(1+{r^2\over r_0^2}\r)^{-1}
\eeq
 with $\s_\O r_0=b^{-1/2}r_c$.  For $b=1.5$ we find $\s_\O r_0=4.5 h^{-1}
\rmn{Mpc}$.  The meaning of $r_0$ will be discussed later, but we
immediately see that for small $r_0$ the model approximates the divergent
galaxy correlation function better.

 From equation (\ref{xiO:xie}) we find
\beq
\xi_\e(r)={1\over\s_\e^2}
\ln\bigg({r_0^2\ee^{\s_\e^2}+r^2\over r_0^2+r^2}\bigg).
\eeq
 From (\ref{Pe:xie}) the power spectrum is
 \begin{eqnarray}
\mod{P_\e(k)}^2
	&=&{1\over2\pi}
	\int_{-\infty}^{+\infty}\d x\,{\ln{{r_0^2\ee^{\s_\e^2}+x^2}
	\over{r_0^2+x^2}}\ee^{\i kx}}\nonumber\\
	&=&{1\over{k}}\l[\exp(-kr_0)-\exp(-kr_0\ee^{\s_\e^2/2})\r].
\end{eqnarray}

The significance of $r_0$ now emerges: it determines how quickly
$\mod{P_\e(k)}$ approaches zero at large $k$. The smaller the value of $r_0$
the larger must be the wavenumber $k_{\rmn{max}}$ up to which we must sum
the discrete Fourier transform from which we obtain realizations of $\O(x)$.
Physically, we should think of $r_0$ as the scale on which the matter
distribution is smoothed by the finite width of our telescope beam and the
diameters of the objects we are looking at.  If we take $r_0=10 h^{-1}
\kpc$, we have $\s_\e^2=12.21$.

Due to computational constraints and limitations on sampling imposed by
Nyquist's theorem, it was impracticable to generate a single random field on
the range $0<z<3$.  Instead, we divided this interval into 100 subintervals
and create a scaled random field on each of them.  This procedure destroys 
correlations between different intervals but these are
physically unimportant because the correlation function is negligible at 
such large distances.

\section{Results and discussion}
\label{sec:discussion}

\begin{figure}
\centerline{\epsfxsize=16pc\epsfbox{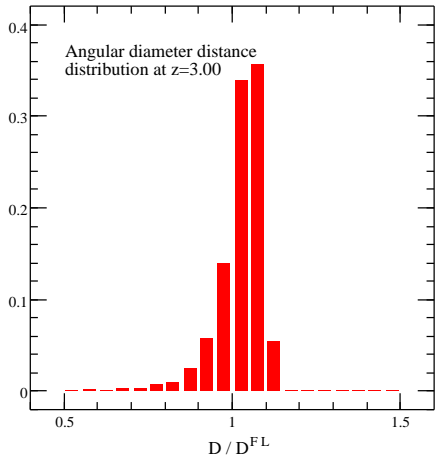}}
\caption{The distribution of angular-diameter distance at $z=3$}
\label{fig:dist}
\end{figure}

\subsection{Distribution of angular-diameter distances}

Once a realization of $\O(x)$ has been constructed, it is straightforward
to solve equation (\ref{new:distance}) for $D$ at any given value of $z$.
We repeated this operation for approximately 4000 realizations of $\O(x)$
to determine the distribution of angular-diameter  distances at $z=3$.
Fig.~\ref{fig:dist} shows this distribution.
The distances are rescaled to the standard \FL\ value
 \beq
D^{\rm FL}={2 \over{1+z}}\l(1-{1\over{\sqrt{1+z}}}\r).
\eeq
 An important point on the graph is the Dyer--Roeder distance corresponding
to an empty light beam:
 \beq
D^{\rm DR}=\fracj25\l(1-{1\over{(1+z)^{5/2}}}\r),
\eeq
 which for $z=3$ gives $D^{\rm DR}/D^{\rm FL}=1.55$.

We see that the distribution is strongly peaked on the Dyer--Roeder side of
$D^{\rm FL}$, with a long tail on the \FL\ side.  This is expected because
regions within which the density is below average occupy the great majority
of the volume of the Universe. Hence, many light paths sample only
low-density regions and the distribution in Fig.~\ref{fig:dist}
is shifted towards $D^{\rm DR}$.  However, when the light beam does encounter
a galaxy or other matter aggregation, it is strongly lensed.  These events
decrease the diameter distance and give rise to the tail on the
\FL\ side.

\begin{figure}
\centerline{\epsfxsize=23pc\epsfbox{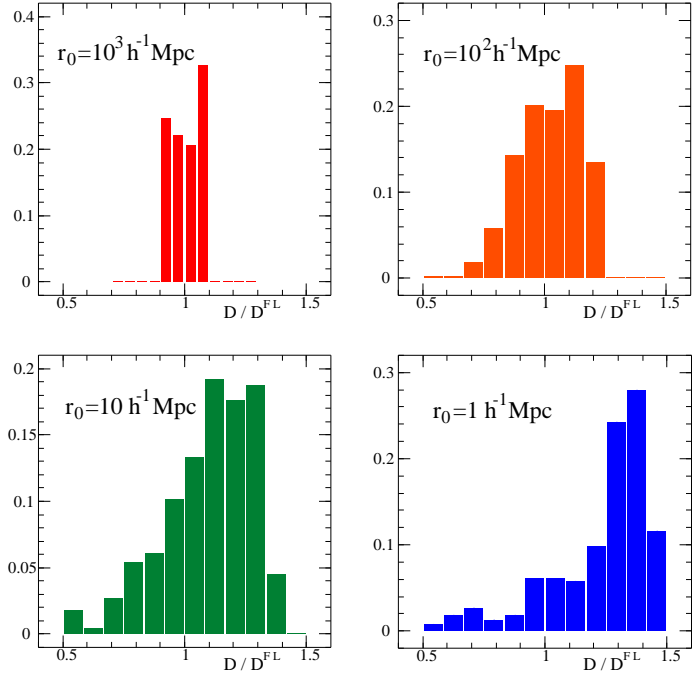}}
\caption{Effect of smoothing on the angular diameter distance distribution}
\label{fig:smooth}
\end{figure}

It is important to understand the impact that smoothing of the matter
distribution has on our results. It is computationally convenient to
investigate this for an unrealistic case: we take the correlation length to
be 100 times its true value. That is, we investigate the case in which
$\s_\O r_0=450 h^{-1} \rmn{Mpc}$.  Fig.~\ref{fig:smooth} shows our results.

For large $r_0$ the matter distribution is rather homogeneous, so the
distribution of $D$ is narrow and peaked near $D^{\rm FL}$. As $r_0$ is
decreased the universe becomes strongly inhomogeneous and the distribution
of $D$ becomes broader. Simultaneously, its peak shifts towards the
empty-beam distance $D^{\rm DR}$.

\subsection{Implications for $q_0$ measurements}

One of the most important undetermined quantities of cosmology is $q_0$, the
deceleration parameter. For a flat universe ($K=0$), the angular-diameter
distance $D$ is related to $q_0$ by
 \beq\label{DRz}
D(q_0,z)={R\over1+z},
\eeq
 where
 \beq\label{Rfromq}
R\equiv\int_{1}^{1+z}{{\d u}\over{\l(\O u^3+(1+q_0-3\O /2)u^2+\fracj12\O
-q_0\r)^{1/2}}}. 
\eeq
 Suppose we attempt to use (\ref{DRz}) to determine $q_0$ from an
observationally determined value of $D(z)$. We assume that the true values
of the cosmic constants are those with  which we have been working: $\O=1$,
$\Lambda=K=0$, and thus that the true value of $q_0$ is
$q_0=\fracj12$. Putting $q_0=\fracj12+\dq$ in equation (\ref{Rfromq}) we find
 \beq
R(\dq)=\int_{1}^{1+z}{{{\d u}\over{u^{3/2}}}\l(1-\dq{{u^2-1}\over{2u^3}}\r)}.
\eeq
 Substituting this value of $R$ into (\ref{DRz}) gives
\beq\label{qerror}
\l.{D\over D^{\rm FL}}\r|_{z=3}=1-0.15\,\dq.\eeq
 This equation relates the error, $\dq$, in the inferred value of $q_0$ to
the ratio of the measured value of $D$ to the value $D^{\rm FL}$ that it
would have if the Universe were homogeneous.  The distribution of $D/D^{\rm
FL}$ shown in Fig.~\ref{fig:dist} is centred on $1.025$ and has spread
$\sim\pm 0.06$. By equation (\ref{qerror}) the error in $q_0$ to which this
gives rise is 
 \beq
\dq=-0.17\pm 0.4.
\eeq
In connection with this result three points should be made:
\begin{itemize}

\item We see that the conventional method of determining $q_0$ from the
angular-diameter redshift relation provides a biased estimator of $q_0$ that
will return significant underestimates of the true value.

\item Even perfect measurements of $D(z)$ will return values of $q_0$ that
are widely scattered. The breadth of this scatter is such that an accurate
determination of $q_0$ would require an extremely
large sample and a sophisticated statistical analysis of the data.

\item The errors in $q_0$ to which inhomogeneities give rise depend  on the
scale of observed objects because this scale determines the effective
spectrum of the inhomogeneities. Larger objects will yield smaller errors.

\end{itemize}

This last point is unfortunate because, as Kellermann (1993) has emphasized,
small objects are much more likely to constitute standard measuring rods
than large objects, such as giant radio sources, whose linear sizes are
likely to be sensitive to the mean cosmic density.

\section{Conclusion}

We have used the theory of gravitational lensing to derive the conventional
relation between angular-diameter distance and redshift in a \FL\ universe.
The value of this derivation is that it is simple and shows that the
tendency of the angular diameter of a distant object to increase with
$\Omega$ arises because rays coming from the object are focused by matter
that lies within the telescope beam. Hence, the angular diameter of an
object is sensitive to the precise disposition of matter in the neighborhood
of the telescope beam: move matter just out of the beam and the apparent
size of the object will diminish.  Equation (\ref{new:distance}) expresses
this fact mathematically.

Since the Universe is strongly inhomogeneous on small scales, telescope
beams to different objects at the same redshift will contain significantly
varying quantities of matter, and the apparent diameters of physically
identical objects at a common redshift will vary. This variation gives rise
to scatter in the angular-diameter distances $D$ of a set of objects that
lie at a common redshift.

We have modelled the distribution of the values of $D$ of objects at redshift
$z=3$ by assuming that the cosmic density field follows a lognormal
distribution that matches the observed clustering of galaxies for bias
parameter $b=1.5$. The distribution of $D$ is very skew, with its peak at a
value that exceeds that associated with the corresponding homogeneous
universe, $D^{\rm FL}$, and a long tail to values smaller than $D^{\rm FL}$.
In consequence of this skewness, the conventional technique for measuring
$q_0$ from measurements of $D(z)$ will systematically underestimate $q_0$.

The width of the distribution of $D$ at given $z$ depends upon the assumed
power spectrum $P(k)$ of the cosmic density field. The true power spectrum
is thought to have considerable power on small scales, and this power will
generate a very broad distribution of $D$ for objects of small angular size.
When the angular diameters of highly extended objects are measured, only
power on scales comparable to or larger than the linear size $r_0$ of the
objects will contribute to the scatter in $D$. Hence such measurements will
yield less scattered values of $D$. For $r_0=10h^{-1}$kpc we estimate that
$D$ will scatter by $\sim\pm6\%$ at redshift $z=3$. Unfortunately, even this
small scatter will cause the derived values of $q_0$ to scatter by as much
as $\pm0.4$. The scatter in values of $q_0$ that are derived from
angular-diameter distances to parsec-sized objects such as those studied by
Kellermann (1993), will be very much larger still.

\end{document}

%% file: mnextra.tex
\font\fivebmi=cmmib6
\font\sixbmi=cmmib6	\skewchar\sixbmi='177
\font\ninebmi=cmmib10 at 9pt 	\skewchar\ninebmi='177
\newfam\bmifam
\textfont\bmifam=\ninebmi
\scriptfont\bmifam=\sixbmi
\scriptscriptfont\bmifam=\fivebmi
\def\bmi{\fam\bmifam\ninebmi}

\mathchardef\alpha="710B
\mathchardef\beta="710C
\mathchardef\gamma="710D
\mathchardef\delta="710E
\mathchardef\epsilon="710F
\mathchardef\zeta="7110
\mathchardef\eta="7111
\mathchardef\theta="7112
\mathchardef\iota="7113
\mathchardef\kappa="7114
\mathchardef\lambda="7115
\mathchardef\mu="7116
\mathchardef\nu="7117
\mathchardef\xi="7118
\mathchardef\pi="7119
\mathchardef\rho="711A
\mathchardef\sigma="711B
\mathchardef\tau="711C
\mathchardef\upsilon="711D
\mathchardef\phi="711E
\mathchardef\chi="711F
\mathchardef\psi="7120
\mathchardef\omega="7121
\mathchardef\varepsilon="7122
\mathchardef\vartheta="7123
\mathchardef\varpi="7124
\mathchardef\varrho="7125
\mathchardef\varsigma="7126
\mathchardef\varphi="7127

\def\chaphead{}
\newcount\eqnumber
\def\today{\ifcase\month\or
 January\or February\or March\or April\or May\or June\or
 July\or August\or September\or October\or November\or December\fi
 \space\number\day, \number\year}

\def\eqnam#1{\xdef#1{(\chaphead\the\eqnumber}}

\def\newe{(\hbox{\chaphead\the\eqnumber})\global\advance\eqnumber by 1}
\def\firste{(\hbox{\chaphead\the\eqnumber a})\global\advance\eqnumber by 1}
\def\laste#1{\advance\eqnumber by -1%
	(\hbox{\chaphead\the\eqnumber #1})\advance\eqnumber by 1}

\def\refe#1{\advance\eqnumber by -#1 (\chaphead\the\eqnumber
     \advance\eqnumber by #1 }

\def\i{\relax\ifmmode{\rm i}\else\char16\fi}
\def\e{{\rm e}}
\def\frac#1#2{{\textstyle{#1\over#2}}}

\def\d{{\rm d}}
\def\dddot#1{\ddot#1\kern-1.4pt\dot{\phantom{#1}}\kern-3pt}

\def\spose#1{\hbox to 0pt{#1\hss}}
\def\s#1{\widetilde{#1}}
\def\=#1{\overline{#1}}

\def\lta{\mathrel{\spose{\lower 3pt\hbox{$\mathchar"218$}}
     \raise 2.0pt\hbox{$\mathchar"13C$}}}
\def\gta{\mathrel{\spose{\lower 3pt\hbox{$\mathchar"218$}}
     \raise 2.0pt\hbox{$\mathchar"13E$}}}

\def\kpc{{\rm\,kpc}}

\def\annrev #1 #2 {ARA\&A, #1, #2}
\def\aa #1 #2 {A\&A, #1, #2}
\def\aasupp #1 #2 {A\&AS, #1, #2}
\def\aj #1 #2 {AJ, #1, #2}
\def\apj #1 #2 {ApJ, #1, #2}
\def\apjlett #1 #2 {ApJ, #1, #2}
\def\apjsupp #1 #2 {ApJS, #1, #2}
\def\ban #1 #2 {Bull.\ Astron.\ Inst.\ Netherlands, #1, #2}
\def\mn #1 #2 {MNRAS, #1, #2}
\def\nature #1 #2 {Nat, #1, #2}
\def\pasj #1 #2 {PASJ, #1, #2}
\def\pasp #1 #2 {PASP, #1, #2}